\documentclass[aps,twocolumn,prl,groupedaddress,showpacs,showkeys,preprintnumbers]{revtex4}

\usepackage[polish,english]{babel}

\usepackage[T1]{fontenc}
\usepackage[latin2]{inputenc}

\newcommand{\be}{\begin{equation}}
\newcommand{\ee}{\end{equation}}
\newcommand{\bea}{\begin{eqnarray}}
\newcommand{\eea}{\end{eqnarray}}

\usepackage{xspace}

\newcommand{\invM}{\ensuremath{{\cal M}}\xspace}

\newcommand{\ep}{\ensuremath{\epsilon}\xspace}
\newcommand{\ie}{\ensuremath{i\epsilon}\xspace}
\newcommand{\la}{\ensuremath{\lambda}\xspace}

\newcommand{\de}{\ensuremath{\delta}\xspace}
\newcommand{\De}{\ensuremath{\Delta}\xspace}
\newcommand{\+}{\ensuremath{\dagger}\xspace}

\newcommand{\f}{\ensuremath{f_\lambda}\xspace}

\newcommand{\F}{\ensuremath{{\cal F}}\xspace}
\newcommand{\cH}[1][]{\ensuremath{{\cal H}_{\lambda #1}}\xspace}

\newcommand{\dx}{d} % \mathrm{d} used in integration measures

\newcommand{\ket}[1]{\ensuremath{\left|#1\right\rangle}}
\newcommand{\bra}[1]{\ensuremath{\left\langle#1\right|}}
\newcommand{\braket}[2]{\ensuremath{\left\langle#1\mid #2\right\rangle}}
\newcommand{\lsub}[2]{\,\vphantom{#2}_{#1}\!{#2}}
\newlength{\mwsuplength}

\newcommand{\centeredgraphics}[2]{\parbox[c]{#1}{\includegraphics[width=#1]{#2}}}

\usepackage{amsmath} %\text etc.
\DeclareSymbolFont{AMSa}{U}{msa}{m}{n}
\DeclareMathSymbol\square           {\mathord}{AMSa}{"03}

\usepackage{graphicx}
\usepackage{color}
\definecolor{mwred}{rgb}{0.5,0,0}
\definecolor{mwgreen}{rgb}{0,0.5,0}
\definecolor{mwblack}{rgb}{0,0,0}
\usepackage[pageanchor=true,hyperindex=true,colorlinks=true,
bookmarks=true,pdfpagemode=None,bookmarksnumbered=true,
bookmarksopen=false,bookmarksopenlevel=0,plainpages=false,
pdfstartview=FitBH,pdffitwindow=true]{hyperref} 
\hypersetup{linkcolor=mwred,citecolor=mwgreen}
\hypersetup{colorlinks=true}

\hypersetup{bookmarksopen=false}

\usepackage{pslatex}

\begin{document}
\preprint{IFT-2005-29}

\title{S-matrix calculus using effective particles in the Fock space}
\author{Marek Więckowski}
\email[]{wiecko@fuw.edu.pl}
\homepage[]{http://www.fuw.edu.pl/~wiecko}

\affiliation{Institute of Theoretical Physics, Warsaw University, 
	ul. Ho\.za 69, 00-681 Warsaw, Poland}

\date{14th November 2005}

\begin{abstract}
This article describes a method for calculating S-matrix elements using 
Hamiltonians obtained in the renormalization group procedure for effective 
particles. It is shown that the scattering amplitudes obtained using a 
canonical Hamiltonian $H^\De$ with counterterms are the same as those obtained 
using a renormalized Hamiltonian for effective particles, \cH.  
The result is independent of the
ultraviolet cutoff \De and the renormalization-group parameter \la.
\end{abstract}
\maketitle

\section{Introduction}
The field-theoretical approach to strong interactions based on the
renormalization group procedure for effective particles (RGPEP) has
developed considerably in recent years.  Most progress has been made in
bound-state problems
\cite{Glazek-Wieckowski,Maslo-phd,Glazek:qqbar,Narebski:2002tj,Mlynik:flambda,Mlynik-acta} 
and the structure of  effective theories \cite{poincare, Ggluon}.
This article deals with
scattering processes \cite{Goldberger-GellMann}
in the language of effective (i.e., constituent) particles.
The issue is important because hadrons are understood in terms of their
constituents and a precise definition of the constituents in quantum
field theory is required for further progress. In particular,
such a definition should be applicable in perturbative scattering theory
and reproduce known results.

This article presents a perturbative description of the
scattering of single physical particles (not their bound states) 
using effective Hamiltonians.
I address two specific questions relating to this description.
The first question concerns the removal of divergences. The standard
perturbative description of scattering uses Feynman diagrams
\cite{Feynman:spacetime-quantum},
based on a formal derivation of expressions for the S matrix.
In the derived expressions one introduces a regularization of each 
divergent loop \cite{tHooft-Veltman} and
constructs counterterms that remove divergences. 
Covariant regularization simplifies the task of finding counterterms because 
one can use symmetries to limit their structure.
Earlier works by
Yan \cite{Yan:1234} followed a similar path within the context of
light-front Hamiltonian theory \cite{Brodsky:1997de}.  In recent works by
Ligterink and Bakker \cite{Ligterink:1994tm} formal covariant
Feynman-diagram expressions are rewritten in a systematic way in terms of
equivalent light-front expressions. See also \cite{Thorn:planar1,Thorn:planar2}
in the context of planar diagrams. 
RGPEP introduces systematically regularized and renormalized 
Hamiltonians that are also applicable in bound-state equations at 
a later stage. 
Hamiltonians are regulated from the very beginning and
counterterm operators are found before one considers a scattering matrix. 
The resulting renormalized theory
may therefore lead to non-divergent non-perturbative predictions.  The
rule for finding the counterterms in the initial canonical Hamiltonian is
that matrix elements of the effective Hamiltonians should not depend on
the ultraviolet cutoff \De, when $\De\to\infty$. Since this way of constructing
counterterms does not refer to any S matrix directly,
we must ask whether the counterterms found using RGPEP
secure a divergence-free S matrix.

The second question concerns the use of creation and annihilation
operators for effective particles. These operators depend on the RGPEP
parameter \la, the inverse of the size of the effective
particles. However, \la is a parameter of a unitary rotation of the Fock-space
basis and, as such, should not influence physical results.
The second question is thus whether the S matrix calculated 
using the effective Hamiltonian \cH is independent of \la.

\section{RGPEP}

In local quantum field theories, canonical Hamiltonians usually lead to 
divergent results. 
It is necessary to introduce an ultraviolet cutoff \De and
construct counterterms in the regulated canonical Hamiltonian $H^\De$ that
remove the dependence of physical results on the regularization. 
RGPEP is based on the
observation that if the Hamiltonian can be re-written as \cH{},
in which every interaction vertex contains  a form
factor \f of width \la{}, then the values of observables predicted by
the Hamiltonian \cH will not depend on \De provided that there is no explicit
\De-dependence in matrix elements of \cH \cite{sim1,sim2}.

In RGPEP a family of unitarily equivalent effective-particle
operators~$a_\la^\+$ is defined for each bare-particle operator
$a_\infty^\+$ \cite{Ggluon}:
\be
a_\la^\+=U_\la a_\infty^\+ U_\la^\+\;. \label{eq:RGPEP:ainf=UalaU}
\ee
Bare particles of the initial canonical theory correspond to $\la=\infty$.  
The same Hamiltonian operator can be expressed in terms of either basis:
\be
H=\cH(a_\la)=H^\De(a_\infty)\;,
\label{H=H=H}
\ee
but with different coefficients. Vertices in \cH contain the form factor
\f. RGPEP equations 
can be solved in perturbation theory, leading to expressions for $U_\la$ and
$\cH$ and allowing the ultraviolet structure of the counterterms in
$H^\De$ to be determined.

\section{S matrix in terms of \texorpdfstring{$H^\De$}{H De}}

The expression for the S matrix in a regularized
light-front Hamiltonian theory can be derived using a procedure 
similar to the textbook derivation of Feynman diagrams
\cite{Bjorken-Drell}. The key steps are summarized below in order to
facilitate further discussion.

It is assumed that the matrix elements of full interacting fields
$\phi_\infty(x)$ (being combinations of bare-particle creation and
annihilation operators $a_\infty^\+$ and $a_\infty$) 
can be approximated in the distant
past by similar matrix elements of free fields $\phi_{in}(x)$
(being combinations of physical-particle creation and
annihilation operators $a_0^\+$ and $a_0$):
\be
\lim_{x^+\to-\infty} \bra{\beta}\phi_\infty(x^\mu)\ket{\alpha}=
\sqrt{Z^\De}
\lim_{x^+\to-\infty}\bra{\beta}\phi_{in}(x^\mu)\ket{\alpha}
\;,
\label{eq:app:phi in = lim phi}
\ee
where $\ket\alpha $ and $\ket\beta$ represent normalized packets of
well-separated particles.  For
massive particles, the above limits for the light-front time
$x^+\to-\infty$ are equivalent to the limits $x^0\to-\infty$.

Using this asymptotic condition, S-matrix elements can be written for all
$p_i\neq q_j$ as:
\begin{multline}
\lsub{out}{\braket{p_1\dots p_{n_1}}{q_1\dots q_{n_2}}}_{in}=
\left(\frac{i}{\sqrt{Z^\De}}\right)^{n_1+n_2} \prod_{i=1}^{n_2}\int \dx^4 x_i
\times\\
\prod_{j=1}^{n_1}\int \dx^4 y_j e^{-iq_{m,i}x_i}
\left(\overrightarrow{\square}_{xi} +m^2\right)
\bra{0}T_{(+)}\Big[
\phi_\infty(y_1)\dots
\\
\dots\phi_\infty(y_{n_1})\phi_\infty(x_1)\dots\phi_\infty(x_{n_2})\Big]
\ket{0}
\left(\overleftarrow \square_{yj}+m^2\right)
e^{ip_{m,j}y_j}
\label{eq:app:LSZ}\;,
\end{multline}
where $T_{(+)}$ denotes ordering in the light-front ``time'' $x^+$, 
$\dx^4 x=\dx^2x^\perp\dx x^-\dx x_-$
and $m$ 
is the physical mass of one particle. 
The momenta in the exponents on the right-hand
side have energy components that fulfill the dispersion relation with
the physical mass.
This is the light-front analogue of the 
Lehmann-Symanzik-Zimmermann (LSZ) formula
\cite{artLSZ}.
When any $p_i$ is equal to any $q_j$, there are
additional forward-scattering terms.

Assuming unitary equivalence of the fields $\phi_\infty$ and $\phi_{in}$:
\be
a_{\infty,\vec k}(x^+)= U^{-1}(x^+)a_{0\vec k}(x^+)U(x^+)\;,
\label{eq:app:a=UaU}
\ee
Eq.~(\ref{eq:app:LSZ}) can be expanded in a perturbative series in powers of
the renormalized canonical interaction Hamiltonian,
\bea
H_I^\De(x^+)&:=&H^\De(a_0)-H_0(a_0)\\
H_0(a_0)&:=&\int[k]\frac{m^2+k^{\perp2}}{k^+}a_{0\vec k}^\+ a_{0\vec k}
\;,
\label{eq:h0:3+1}
\eea
namely:
\begin{multline}
\bra{0}T_{(+)}\left[
\phi_\infty(x_1)\dots\phi_\infty(x_n)\right]\ket{0}=
\bra{0}T_{(+)}\Big[
\phi_{in}(x_1)\dots 
\\
\dots\phi_{in}(x_n)
\exp \left(-i\int_{+\infty}^{-\infty} 
H_I^\De(x^+)\cdot\frac{1}{2}\dx x^+\right)
\Big]\ket{0}\;. \label{eq:app:tau(H)}
\end{multline}
The above steps apply in the presence of
regulators in the light-front Hamiltonians.

\section{S matrix in terms of \texorpdfstring{\cH}{H\_lambda}}
Instead of the field $\phi_\infty(x)$ used above,
we may introduce an effective field $\phi_\la(x)$ to represent
the same physical situation. 
$\phi_\la(x)$ is defined using the
effective-particle creation operators $a_\la^\+$ as the Fourier coefficients.
The evolution of both $\phi_\infty(x)$ and $\phi_\la(x)$
is determined by the same evolution operator~$H$ of Eq.~(\ref{H=H=H}), but
their matrix elements have different asymptotic behaviors. Instead of
(\ref{eq:app:phi in = lim phi}), for $\phi_\la$ we have:
\be
\lim_{x^+\to-\infty}{\bra{\beta}}\phi_\la(x^\mu)\ket{\alpha}
=
\lim_{x^+\to-\infty}\sqrt{Z_\la}\;\;
{\bra{\beta}}\phi_{in}(x^\mu)\ket{\alpha}\;.
\label{eq:app:asymtotic:effective}
\ee
The derivation of the LSZ formula for the same physical S-matrix element can
be repeated using $\phi_\la(x)$, leading to:
\begin{multline}
\lsub{out}{\braket{p_1\dots p_{n_1}}{q_1\dots q_{n_2}}}_{in}=
\left(\frac{i}{\sqrt{Z_\la\vphantom{{}^\De}}}\right)^{n_1+n_2} \prod_{i=1}^{n_2}\int 
\dx^4 x_i \times\\
\prod_{j=1}^{n_1}\int \dx^4 y_j
e^{-iq_{m,i}x_i}
\left(\overrightarrow{\square}_{xi} +m^2\right)
\bra{0}T_{(+)}\Big[
\phi_\la(y_1)\dots
\\
\dots\phi_\la(y_{n_1})
\phi_\la(x_1)\dots\phi_\la(x_{n_2})\Big]
\ket{0}
\left(\overleftarrow \square_{yj}+m^2\right)
e^{ip_{m,j}y_j}\;.
\label{eq:app:LSZ:effective}
\end{multline}

Substituting Eq.~(\ref{eq:RGPEP:ainf=UalaU}) into
(\ref{eq:app:a=UaU}), we get:
\be
a_{\vec k,\la}(x^+)= W_\la^{-1}(x^+)a_{0\vec k}(x^+)W_\la(x^+)\;,
\label{eq:ala=WaW}
\ee
where
\be
W_\la(x^+)=U(x^+) U_\la^\+(x^+)\;,
\ee
and $U_\la(0)=U_\la$ from Eq.~(\ref{eq:RGPEP:ainf=UalaU}).
Eq.~(\ref{eq:ala=WaW}) is an analogue of (\ref{eq:app:a=UaU})  for
effective-particle operators~$a_\la$.

We may now repeat the steps from 
Eqs.~(\ref{eq:app:a=UaU})-(\ref{eq:app:tau(H)}) using $W_\la(x^+)$ instead of
$U(x^+)$. This leads to a similar perturbative expansion of the S
matrix:
\begin{multline}
\bra{0}T_{(+)}\left[\phi_\la(x_1)\dots\phi_\la(x_n)\right]\ket{0}
=\bra{0}T_{(+)}\Big[
\phi_{in}(x_1)\dots
\\
\dots \phi_{in}(x_n)
\exp \left(-i\int_{-\infty}^{+\infty} 
\cH[,I](x^+)\cdot\frac{1}{2}\dx x^+\right)
\Big]\ket{0}\;, \label{eq:app:tau(Hla)}
\end{multline}
but with an interaction Hamiltonian defined as follows:
\be
\cH[,I]=\cH(a_0)-H_0(a_0)\;.
\ee

The results can be summarized in the following theorem:\\
{\it
The same S matrix describing the scattering of physical particles can be
obtained using either:
\begin{enumerate}
\item[(i)] A bare Hamiltonian $H^\De$ and representing the incoming/outgoing
particles by bare-particle creation and annihilation operators $a_\infty^\+$
and  $a_\infty$; or
\item[(ii)] An effective Hamiltonian \cH and creation and annihilation
operators for effective particles,
$a_\la^\+$ and $a_\la$.
\end{enumerate}
In each order of perturbation theory the result for the S matrix is the
same, provided that the unitary relation between $a_\infty$ and $a_\la$
(i.e., between $H^\De$ and \cH) is fulfilled up to this order. }

For (i), the S matrix is obtained using the LSZ formula
(\ref{eq:app:LSZ}) with wave-function renormalization factors $Z^\De$ and
perturbative expansion (\ref{eq:app:tau(H)}) in powers of the bare
interaction Hamiltonian $H^\De_I$. For (ii), the S matrix is
obtained using LSZ formula (\ref{eq:app:LSZ:effective}) with different
wave-function renormalization factors $Z_\la$ and a perturbative
expansion (\ref{eq:app:tau(Hla)}) in powers of the effective interaction
Hamiltonian \cH[,I].

\section{Example: tree amplitude in a scalar model}
%\subsection{Bare Hamiltonian}
In order to illustrate the meaning of the preceding discussion
using the simplest possible example,
I consider here a Hamiltonian describing the
interaction of three bosonic fields in 1+1 space-time dimensions:
\bea
H&=&H_{0}+H_{e}\\
H_{0}&=&
\int[k] \frac{m^2}{k^+} \left(
a_{\vec k,e}^\+ a_{\vec k,e}+
e_{\vec k,e}^\+ e_{\vec k,e}+
q_{\vec k,e}^\+ q_{\vec k,e}\right)\;,
\label{eq:H0:1+1}
\eea
\begin{multline}
H_{Y}=\int[k_1k_2k_3]4\pi\de(k_1^++k_2^+-k_3^+)
\Big(\frac{e_q}{2}q_{k1}^\+q_{k2}^\+a_{k3}+\\
\frac{e_e}{2}a_{k3}^\+e_{k1}e_{k2}+H.c.\Big)\;=
\centeredgraphics{0.9cm}{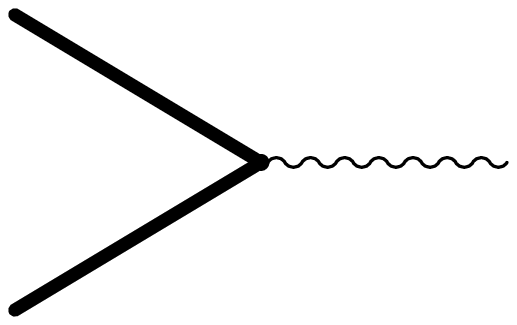}+
\centeredgraphics{0.9cm}{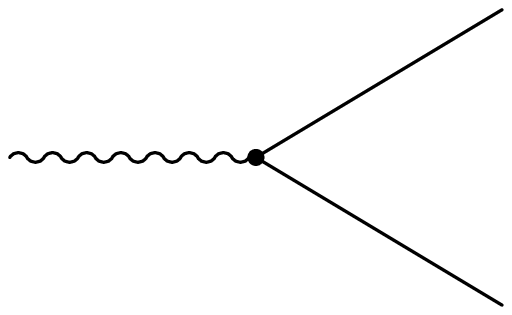}+H.c.\;,\label{eq:HY:s}
\end{multline}
where $[k]=dk^+\theta(k^+)/\left(4\pi k^+\right)$ and
$\left[a_k,a_p^\+\right]=4\pi k^+\de(k^+-p^+)$.  Since this model is not
divergent, the bare Hamiltonian does not require regularization and
renormalization, and the comparison between the bare and effective
descriptions is simple. Nevertheless, the Hamiltonian contains vertices
of a structure 
similar to those present in realistic quantum field theories.

%\subsection{Effective Hamiltonian up to order 
%\texorpdfstring{$e^2$}{e\texttwosuperior}}
The effective Hamiltonian \cH is calculated using the RGPEP differential 
equations \cite{Ggluon}.
In the zeroth order in powers of the charges $e_q$ or $e_e$:
$\cH[0]=H_{0}$.

In the first order, the effective Hamiltonian $\cH^{(e)}$ is simply
the sum of all bare vertices with form factors $\f$:
\bea
\cH[>\!\!-]&=&\f\centeredgraphics{0.9cm}{can-e-p2qq}\;,\\
\cH[-\!\!<]&=&\f\centeredgraphics{0.9cm}{can-e-ee2p}\;,
\eea
(with $q^\+_\la q^\+_\la a_\la $ and $a^\+_\la e_\la e_\la $ operator
structure,  using the same conventions as in Eq.~(\ref{eq:HY:s})) 
plus their Hermitian conjugates.

In the second-order, the only part of the effective Hamiltonian that 
contributes to the calculation below
is the one with $q^\+_\la q^\+_\la e_\la e_\la$,
represented by the diagram:
\be
\cH[>\!\!-\!\!<]^{(e^2)} = f_{ac} \F^{(2)}_{abc}
\centeredgraphics{1.2cm}{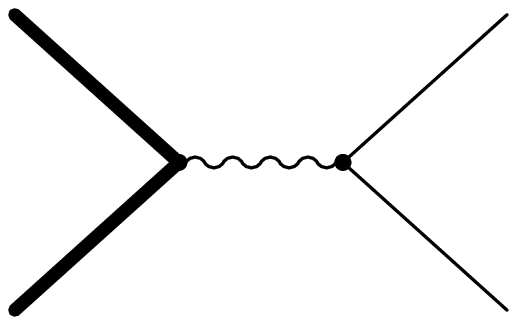}\;.
\label{eq:example:H2}
\ee
In this expression
$f_{ac}=\exp(-ac^2/\la^2)$ and
$\F _{abc}^{(2)}  =  \left(P_{ba}^{+}ba+P_{bc}^{+}bc\right)/\left(ba^{2}+bc^{2}
\right)\cdot\left(f_{ba}f_{bc}-1\right)$.
$a$, $b$ and $c$ mark the left-most, intermediate, and right-most
configurations of particles in interactions;
combinations of two letters denote the differences of the squares of free
invariant masses, e.g., $ab=\invM_{a}^2-\invM_{b}^2$
(see \cite{Ggluon} for details of this notation).

%\subsection{Scattering amplitude calculated using 
%\texorpdfstring{$H^\De$}{H\^{}Delta} and 
%\texorpdfstring{$\cH$}{H\_lambda}}

In the order $e^2$, $H^\De$ leads to the following S matrix for
$ee\to qq$ scattering:
\be
H_{>\!\!-}\frac{1}{P^-_0-H_0+\ie} H_{-\!\!<}=
\frac{P^+_{ab}}{ab+\ie} \centeredgraphics{1.2cm}{S--e2-empty}
\label{eq:tree-eeTOhadr}\;.
\ee

The same result is found when the
effective Hamiltonian~\cH is used. There are two contributions: one is from
$\cH^{(e)}$ acting twice:
\be
\cH[>\!\!-]^{(e)}\frac{1}{P^-_0-H_0+\ie}\cH[-\!\!<]^{(e)}=
\frac{P^+_{ab}}{ab+\ie} f_{ab}^2
\centeredgraphics{1.2cm}{S--e2-empty}
\label{eq:example:part-of-S-from-HlaHla}
\;,
\ee
and the other is from $\cH[>\!\!-\!\!<]^{(e^2)}$. When
$\cH[>\!\!-\!\!<]^{(e^2)}$ contributes to the S matrix, due to the energy
conservation  $f_{ac}\equiv 1$ and $\F^{(2)}_{abc}$ simplifies to:
\be
\F^{(2)}_{aba}=\frac{P^+_{ba}}{ba}(f_{ab}^2-1)\;.
\ee
Thus on the energy shell:
\be
\cH[>\!\!-\!\!<]^{(e^2)}\big|_{ac=0}= (1-f_{ab}^2)\frac{1}{P^-_0-H_0} 
H_{>\!\!-}H_{-\!\!<}\;.
\label{eq:example:part-of-S-from-H2la}
\ee
For momenta distant from the pole,
(\ref{eq:example:part-of-S-from-HlaHla})  and the $f_{ab}^2$ term in
(\ref{eq:example:part-of-S-from-H2la}) cancel and the remainder reproduces the
S matrix obtained using the bare Hamiltonian, (\ref{eq:tree-eeTOhadr}).
For momenta close to the pole, $ab$ goes to zero (and $f_{ab}\to1$)
and the whole contribution to the pole comes from
(\ref{eq:example:part-of-S-from-HlaHla}); the result for the residue
in the pole is the same as that calculated using the bare
Hamiltonian~(\ref{eq:tree-eeTOhadr}). The analytic structures
of the amplitudes obtained using $H^\De$ and $\cH$ are thus also the same 
when $\ep\to0$.

This example also demonstrates the key difference between the general
off-shell Hamiltonian term, e.g., (\ref{eq:example:H2}), and the
corresponding on-shell S-matrix element, e.g.,
(\ref{eq:example:part-of-S-from-H2la}): 
where $H$ is given, the S matrix can be
calculated, while the S matrix alone is not sufficient to define a
corresponding operator $H$.  This difference is particularly important 
for diverging theories, where perturbative S-matrix renormalization is
of little help in constructing well-defined non-perturbative bound-state
equations.

\section{Consequences of the theorem}
The counterterms found using RGPEP lead to a divergence-free
S matrix.
When we calculate the scattering amplitude using \cH, the results do
not depend on $\De$ in the corresponding order of perturbation theory, when
$\De\to\infty$, owing to the form factors in the Hamiltonian's interaction 
terms. The theorem states that both \cH and $H^\De$ produce the same S-matrix
elements. Thus the scattering amplitude can be obtained using the renormalized
canonical Hamiltonian $H^\De$ and the RGPEP counterterms in $H^\De$ do indeed 
lead to results which are not divergent. This observation is not trivial, 
as the counterterms were found from conditions not directly related to
the S-matrix formalism and the
ultraviolet cutoff dependence of the S matrix could originate from several 
different
sources. For example, the factor $Z^\De$, appearing both in the LSZ
formula (\ref{eq:app:LSZ}) and in the full propagator in realistic theories is a divergent
function of $\De$; the theorem, however, implies that \De-dependence will cancel.

The S matrix calculated using the effective Hamiltonian \cH is independent of
\la. 
If we calculate a
scattering amplitude using $H^\De$, the result does not depend on \la,
since there is no such parameter in $H^\De$. However, the same result can
be obtained using \cH. This means that the effective Hamiltonian \cH leads
to \la-independent results for the S matrix in a given order of
perturbation theory in an appropriately defined coupling constant $e_\la$.
The dependence of $e_\la$ on \la is calculated to the same order.  Again, 
without the theorem this is not obvious: the wave-function
renormalization factors for effective particles, $Z_\la$, depend on $\la$
and there are many terms in the effective Hamiltonian that do not appear
in the canonical Hamiltonian.

\section{Conclusions}

In this article I have examined the applicability of RGPEP to S-matrix
calculations.  I have shown that both a renormalized canonical Hamiltonian
$H^\De$ and the corresponding effective Hamiltonian $\cH$ give the same S
matrix. In the calculation using $H^\De$, physical states of colliding
particles are expressed in terms of bare particles and their interactions
are contained in the renormalized canonical
Hamiltonian $H^\De_I$. In the calculation using \cH,
the physical states are expressed in terms of effective particles 
and the effective interaction Hamiltonian
\cH[,I] with form factors in all vertices is used. 
Since it is known that the effective-particle approach applies in
the case of bound states \cite{Glazek-Wieckowski}, I conclude that
a single approach based on RGPEP is applicable in the description of 
both scattering and bound states of effective, constituent particles. 

A corollary of the
equivalence of the canonical and effective S-matrix calculus is that RGPEP
counterterms remove divergences from the S matrix. 
(How this takes place in detail, at least for low
orders, is an area for future investigation.)
However, RGPEP does not give the finite parts of the counterterms --
these are constrained by the requirement of covariance
\cite{Maslowski-Wieckowski, Glazek:hybrydy, Narebski-hybrydy} (see also
\cite{longpaper,PPR,Burkardt:1991tj}).

A second corollary of the theorem is that the S matrix calculated using \cH is
independent of \la. The example above 
exhibits this feature in a straightforward calculation, since
the overall form factor $f_{ac}$ is equal to~$1$
(due to energy conservation). In
higher orders, the calculation is less simple:
the corresponding form factor \f is no longer equal to $1$, as
\f is defined using free energies, whereas energy denominators and
energy conservation in S-matrix calculations
correspond to the physical masses of colliding
particles.

I do not analyze here the conceptual and technical difficulties
associated with the so-called small-x singularities in gauge theories
(see
\cite{Susskind:1994wr,Wilson:1994gn,Wilson:1994fw,Rozowsky:2000gy}).
However, the general theorem I present also holds where there are
cutoffs on small-plus-momentum fractions.
It is likely that the theorem is also valid in
gauge theories, although this requires verification.

%\bibliography{art}
%\bibliography{art-eprint}
% add: `` See e.g. '' to Bjorken
% remove coma from `` , \bibinfo{note}{and references ''
% bind Feynman et.al. in 1 citation
% bind LSZ in 1 citation

\end{document}